\documentclass[twocolumn, pra, showpacs,superscriptaddress]{revtex4}
\usepackage{graphicx}
\usepackage{amsmath}
\usepackage{bm}
\usepackage{float}
\usepackage{color}
\usepackage{dcolumn}

\begin{document}
\title{Quantum entangled ground states of two spinor Bose-Einstein condensates}
\author{Z. F. Xu}
\affiliation{State Key Laboratory of Low Dimensional Quantum Physics,
Department of Physics, Tsinghua University, Beijing 100084, China}
\author{R. L\"u}
\affiliation{State Key Laboratory of Low Dimensional Quantum Physics,
Department of Physics, Tsinghua University, Beijing 100084, China}
\author{L. You}
\affiliation{State Key Laboratory of Low Dimensional Quantum Physics,
Department of Physics, Tsinghua University, Beijing 100084, China}

\date{\today}

\begin{abstract}
We revisit in detail the non-mean-field ground-state phase diagram
for a binary mixture of spin-1 Bose-Einstein condensates including
quantum fluctuations. The non-commuting terms in the spin-dependent
Hamiltonian under single spatial mode approximation make it
difficult to obtain exact eigenstates. Utilizing the spin
z-component conservation and the total spin angular momentum
conservation, we numerically derive the information of the building
blocks and evaluate von Neumann entropy to quantify the ground
states. The mean-field phase boundaries are found to remain
largely intact, yet the ground states show fragmented and entangled
behaviors within large parameter spaces of interspecies
spin-exchange and singlet-pairing interactions.
\end{abstract}



\maketitle

\section{Introduction}

Ultracold atomic quantum gases with spin degrees of freedom
provide exceptionally clean and idealized testing beds
for studying quantum magnetism \cite{trotzky2008,weld2011}.
Optical trapping from ac Stark shifts of off-resonant
laser fields are capable of equal confinement for all atomic pseudo-spin
components which facilitates research into exciting spinor physics
with atomic quantum gases. As is often employed in studying a trapped Bose gas,
when treating the condensate, we first take a mean-field (MF)
approximation assuming only one eigenvalue of
the single-particle density matrix is macroscopic,
being of order $N$. Second quantization is then limited to the
condensate mode. Such a simple scenario
already allows for many interesting quantum many body phenomena\cite{ueda2010}.

Two popular atomic species often employed in experimental research
on spinor Bose-Einstein condensates (BECs)
\cite{ueda2010,ho1998,ohmi1998,law1998,koashi2000,ciobanu2000,ueda2002},
are $^{87}$Rb and $^{23}$Na atoms. Within each species, their
interactions are dominated by the density-dependent interaction in
comparison to the much weaker spin-dependent interactions. As a
result, single-spatial mode approximation (SMA), whereby the spatial
dependence of the condensate wave function is determined
independent of the spin degrees of freedom, was introduced
\cite{law1998} and remains reasonable as long as the number of atoms
is not too large \cite{yi2002}. Within the MF approximation, the
ground state of a spinor Bose-Einstein condensate (BEC) is found
to be ferromagnetic, polar, or cyclic phases, {\it etc}, determined
by the spin-dependent interactions and the total (hyperfine-)spin
$F$ of the atom. Further theoretical work armed with full quantum
calculations reveal interesting many-body states
\cite{law1998,koashi2000,ueda2002,ho2000}, beyond the scope of those
from MF approximations. For spin-1 condensates, the exact
eigenstates will contain paired spin singlets
\cite{law1998,koashi2000,ho2000}, which become more complex for
higher spin condensates. For example, the spin-2 case involves spin
singlets which can be formed by either two or three atoms
\cite{koashi2000,ueda2002}. A general procedure exists for more
detailed information of the building blocks of eigenstates
determined by their associated generating functions \cite{ho2000b}.
More generally, we can always resort to the means of numerics to
diagonalize the ground-state single-particle density matrix, which
then reveals fragmented ground states if more
than one eigenvalues are being of order $N$\cite{mueller2006}.

Several groups have recently studied spinor condensate mixtures
\cite{luo2007,xu2009,xu2010,zhang2010,shi2010,xu2010b, zhang2011,shi2011},
which also display non-MF features, such as anomalous
quantum fluctuations for each spin components and quantum entangled ground states.
This is first discovered in spinor condensates with more than one orbitals,
for instance, the case of pseudo spin-1/2 condensates \cite{kuklov2002,ashhab2003,shi2006}
for which Kuklov and Svistunov \cite{kuklov2002}
predicted that in the ground states all atoms will have to condense into
two orthogonal spatial orbitals or more due to the conservation of the total spin.
This could result in a condensate ground state being a maximally entangled many-body state.
Shi {\it et al.} replaced the two orbitals with two different atomic species,
a ground state with entangled order parameter follows \cite{shi2006}.
Under MF approximation, we have previously elaborated the ground-state
phase diagram for a condensate mixture of two spin-1 BECs \cite{xu2009}.
The interesting phases are named appropriately
as FF, AA, PP, CC, and MM phases, distinguishing different structures and
interaction parameter spaces.
Furthermore, we provide many beyond MF results based on a
full quantum spin-dependent Hamiltonian \cite{xu2010},
which contains non-commuting terms, forbidding a simple derivation of
the exact eigenstates. For two special cases, commutations are
restored among the generally non-commuting terms.
First, when the interspecies singlet pairing interaction is ignored ($\gamma=0$),
the Hamiltonian is then simply composed of three
operators which obey the angular momentum algebra.
Making use of the eigenstates of single spin-1 condensates,
we directly construct the eigenstates of a binary spin-1 mixture
using the angular momentum coupling representation.
Second, when the interspecies anti-ferromagnetic spin-exchange interaction is strong enough,
the ground state will be forced to develop entanglement between the two species \cite{xu2010},
a result consistent with what is discovered in a spin-1 condensate
placed inside a double well \cite{jack2005}.
Other interesting features are discussed
for $\gamma=0$ revealing fragmentation and quantum entanglement \cite{zhang2010,shi2010,shi2011}.

\section{the model Hamiltonian}

In this revisit we hope to understand quantum entanglement between
two spin-1 condensates when both interspecies
spin-exchange and singlet-pairing interactions are present.
Our study is based on the same model system of
a binary mixture of spin-1 condensates confined in optical traps.
The corresponding field operators that annihilates a boson of species 1 and
species 2 at position $\mathbf{r}$ are described respectively
by $\hat{\Psi}_{M_F}(\mathbf{r})$ and $\hat{\Phi}_{M_F}(\mathbf{r})$,
where $M_F=-1,0,1$ denoting the three Zeeman hyperfine states.
The SMA is adopted for each of the two species,
employing two spatial mode functions $\psi(\mathbf{r})$ and
$\phi(\mathbf{r})$ respectively, and the field operators are
expanded as $\hat{\Psi}_{M_F}(\mathbf{r})=\hat{a}_{M_F}\psi(\mathbf{r})$
and $\hat{\Phi}_{M_F}(\mathbf{r})=\hat{b}_{M_F}\phi(\mathbf{r})$,
with $\hat{a}_{M_F}$ and $\hat{b}_{M_F}$ respectively the
annihilation operators for an atom in the  spin component $M_F$.
In the absence of external magnetic field, the spin-dependent
Hamiltonian for a binary mixture of spin-1 condensates
then becomes the following
\begin{eqnarray}
  \hat{H}_s&=&\frac{1}{2}C_1\beta_1\left(\hat{L}_1^2-2\hat{N}_1\right)
  +\frac{1}{2}C_2\beta_2(\hat{L}_2^2-2\hat{N}_2)\nonumber\\
  &&+\frac{1}{2}C_{12}\beta\hat{L}_1\cdot\hat{L}_2
  +\frac{1}{6}C_{12}\gamma\hat{\Theta}_{12}^{\dag}\hat{\Theta}_{12},
  \label{hamiltonian}
\end{eqnarray}
under the SMA \cite{xu2010}.
The interaction coefficients are $C_1=\int d\mathbf{r}|\psi(\mathbf{r})|^4$,
$C_2=\int d\mathbf{r}|\phi(\mathbf{r})|^4$, and
$C_{12}=\int d\mathbf{r}|\psi(\mathbf{r})|^2|\phi(\mathbf{r})|^2$.
$\beta_1$ ($\beta_2$) is intra-species spin-exchange interaction
parameter of species 1 (2). $\beta$ and $\gamma$ denote
inter-species spin-exchange and singlet-pairing interaction parameters,
respectively. The singlet pairing operator becomes
$\hat{\Theta}^{\dag}_{12}=\hat{a}^{\dag}_1\hat{b}^{\dag}_{-1}
-\hat{a}_0^{\dag}\hat{b}_{0}^{\dag}+\hat{a}^{\dag}_{-1}\hat{b}^{\dag}_1$,
and two angular momentum like operators
$\hat{\mathbf{L}}_1=\sum_{ij}\hat{a}^{\dag}_{i}\mathbf{F}_{ij}\hat{a}_j$
and $\hat{\mathbf{L}}_2=\sum_{ij}\hat{b}^{\dag}_{i}\mathbf{F}_{ij}\hat{b}_j$
obey the usual angular momentum algebra \cite{law1998,wu1996}.
They commute
with atom number operators $\hat{N}_1=\sum_i\hat{a}_i^{\dag}\hat{a}_i$ and
$\hat{N}_2=\sum_i\hat{b}^{\dag}_i\hat{b}_i$. In the above,
$\mathbf{F}_{ij}$ denotes the $(i,j)$ component of the spin-1 matrix $\mathbf{F}$.

\section{Ground-state phase diagram}

As presented in the earlier study of \cite{xu2010}, for the spin-dependent
Hamiltonian of Eq.~(\ref{hamiltonian}),
the first three terms commute with each other,
but they do not commute with the fourth term.
This shows the ground state determined will depend on the interaction parameters.
As a result, we resorted to the special cases
of no interspecies singlet-pairing interaction ($\gamma=0$)
and $C_1\beta_1=C_2\beta_2=C_{12}\beta/2$ \cite{xu2010}.
For the first case of $\gamma=0$, it has already attracted much attention
due to the appearance of fragmented ground states and the associated
entanglement between two species and exotic atomic number fluctuations
\cite{xu2010,zhang2010,shi2010}.

In this study, we will discuss the general case of the full spin-dependent
Hamiltonian of Eq.~(\ref{hamiltonian}).
Whenever the ground state depends on the interaction parameters,
we have to perform a full quantum calculation numerically,
usually this amounts to a full exact
numerical diagonalization for each atom numbers $N_1$ and $N_2$.
Before discussing the numerical results, we want to point out
that there still exist two conserved quantities: the total spin angular momentum and its
$z$-component, as $\hat{L}^2=(\hat{L}_1+\hat{L}_2)^2$ commutes with the spin-dependent
Hamiltonian. As a result, we can elaborate spin structures from building
blocks derived by generating function of the maximum spin states $|l,l_z=l\rangle$,
where we have used quantum numbers $l$ and $l_z$ to denote the common eigenstates of
the angular momentum operators $\hat{L}^2$ and $\hat{L}_z$.

We recall the suitable generating function $G_g(x,y,z)$ for a binary mixture of
two spin-1 condensates derived earlier in Ref. \cite{xu2010}.
From this generating function, we have figured out all six building
blocks for constructing the eigenstate $|l,l\rangle$, which is given by
\begin{eqnarray}
  |l,l\rangle=\sum \mathcal{C}(\{u_i\},\{v_i\},\{w_i\})
  \big(\hat{A}_1^{(1)\dag}\big)^{u_1}
  \big(\hat{A}_0^{(2)\dag}\big)^{u_2}
  \big(\hat{B}_1^{(1)\dag}\big)^{v_1}
  \nonumber\\
  \times
  \big(\hat{B}_0^{(2)\dag}\big)^{v_2}
  \big(\hat{\Gamma}_0^{(1,1)\dag}\big)^{w_1}
  \big(\hat{\Gamma}_1^{(1,1)\dag}\big)^{w_2}|\rm vac\rangle,
  \qquad
  \qquad
  \label{llstate}
\end{eqnarray}
where the six building blocks are
\begin{eqnarray}
  \hat{A}^{(1)\dag}_1&=&\hat{a}_1^{\dag},\nonumber\\
  \hat{A}^{(2)\dag}_0&=&\hat{a}_0^{\dag 2}-2\hat{a}_1^{\dag}\hat{a}_{-1}^{\dag},\nonumber\\
  \hat{B}^{(1)\dag}_1&=&\hat{b}_1^{\dag},\nonumber\\
  \hat{B}^{(2)\dag}_0&=&\hat{b}_0^{\dag 2}-2\hat{b}_1^{\dag}\hat{b}_{-1}^{\dag},\nonumber\\
  \hat{\Gamma}^{(1,1)\dag}_0&=&\hat{\Theta}_{12}^{\dag},\nonumber\\
  \hat{\Gamma}^{(1,1)\dag}_1&=&\frac{1}{\sqrt{2}}\left(\hat{a}^{\dag}_1\hat{b}^{\dag}_{0}
  -\hat{a}^{\dag}_{0}\hat{b}^{\dag}_1\right),
  \label{buildingblocks}
\end{eqnarray}
and $u_i$ ,$v_i$, and $w_i$ satisfy the constrains
\begin{eqnarray}
  u_1+2u_2+w_1+w_2&=&N_1,\nonumber\\
  v_1+2v_2+w_1+w_2&=&N_2,\\
  u_1+v_1+w_2&=&l,\nonumber
  \label{constrains}
\end{eqnarray}
and additionally $w_2=0,1$.
Spin states $|l,l_z\ne l\rangle$ of other magnetization
can be constructed by simply applying $(\hat{L}_{-}^{l-l_z}$
on the state $|l,l\rangle$ as $\hat{L}_-^{l-l_z}|l,l\rangle\propto |l,l_z\ne l\rangle$ (un-normalized).

In most numerical studies, we assume each species contain 100 atoms
$(N_1=N_2=N=100)$. Due to the SO(3) symmetry of our model, we
restrict the Hilbert space into the subspace of zero magnetization
with $l_z=0$ \cite{numerical}. In Fig. \ref{fig1}, we summarize the
extensive numerical results. Since the total spin angular momentum
is conserved, we can use the eigenvalue of the operator $\hat{L}^2$
to distinguish different phase, which is then accompanied by the
information of the building blocks. A total of three constrains
exist for the allowed values of $u_i$, $v_i$, and $w_i$ $(i=1,2)$.
Only three are needed for a solution, we choose the three as $u_2$,
$v_2$ and $w_1$, which are determined numerically from evaluating
the associated expectation values of
$\hat{A}_0^{(2)\dag}\hat{A}_0^{(2)}$,
$\hat{B}_0^{(2)\dag}\hat{B}_0^{(2)}$, and
$\hat{\Gamma}_0^{(1,1)\dag}\hat{\Gamma}_0^{(1,1)}$, respectively.

From extensive numerical results
we construct the ground-state phase diagram as shown in Fig. \ref{fig1}.
Perhaps not surprisingly, it is almost the
same as the MF approximation studied in Ref. \cite{xu2009}.
Each phase is then labeled the same as before \cite{xu2009},
albeit that the meanings can be different due to the
non-commuting operators in the spin-dependent Hamiltonian Eq.~(\ref{hamiltonian}).
In Tab.~\ref{tabl}, we summarize the properties for the four special phases: FF, AA, PP, and CC.
The remaining MM phase still denotes the
phase whose parameters evolve continuously across the phase boundaries.

\begin{table}[H]
\centering
\begin{tabular}[b]{c|c|c|c|c}
\hline\hline
& $\langle\hat{L}^2\rangle$ & $\langle\hat{A}_0^{(2)\dag}\hat{A}_0^{(2)}\rangle$ &
$\langle\hat{B}_0^{(2)\dag}\hat{B}_0^{(2)}\rangle$
& $\langle\hat{\Gamma}_{0}^{(1,1)\dag}\hat{\Gamma}_{0}^{(1,1)}\rangle$ \\ \hline
FF&$ =2N(2N+1)$& $=0$ &$=0$ & $=0$\\ \hline
AA& $=0$ & $\sim 0$ & $\sim 0$ & $\sim N(N+2)$ \\ \hline
PP& $=0$ & $\sim N(N+1)$ & $\sim N(N+1)$ & $\sim0$ \\ \hline
CC1& $=N(N+1)$ & $\sim N(N+1)$& $\sim 0$ & $\sim N(N+2)$ \\ \hline
CC2& $=0 $& $\sim N(N+1) $ & $\sim N(N+1)$ & $\sim N(N+2)$\\ \hline
CC3& $=N(N+1)$ & $\sim 0$ & $\sim N(N+1)$ & $\sim N(N+2)$ \\ \hline\hline
\end{tabular}
\caption{The expectation values for the special operators
in the ground state within different phases.}
\label{tabl}
\end{table}

\section{Entangled ground states}

\begin{figure}[tpb]
\centering
\includegraphics[width=3.2in]{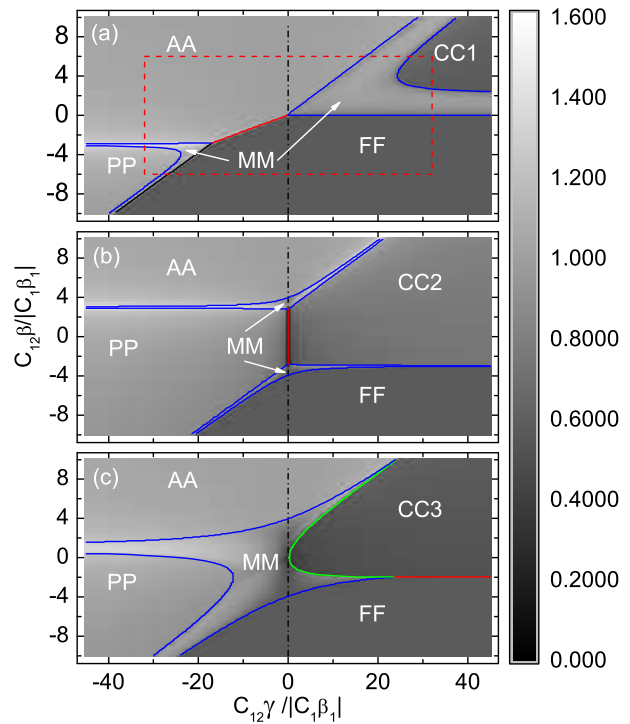}
\caption{(Color online). The ground-state phase diagram and the
corresponding von Neumann entropy distribution at fixed values of
$C_1\beta_1$ and $C_2\beta_2$. Blue solid lines denote continuous
phase transition boundaries. Red solid lines denote discontinuous
phase transition boundaries between two phases with fully determined
total spin angular momentum $l$. The green solid line denotes the
discontinuous phase transition boundary between the phase CC3 and
the MM phase. The black dash-dotted lines correspond to
$C_{12}\gamma=0$, which serve as guides for the eye. The von Neumann
entropy of the ground states are painted by gray scale density
plots, where black (white) color refer to low (high) entanglement
respectively. The three subplots denote fixed intra-specie spin
exchange interaction parameters of $(C_1\beta_1,
C_2\beta_2)/|C_1\beta_1|$ $=$: (a) ($-1$,$-2$) ; (b) ($1,2$); and
(c) ($-1,2$). The red dashed lines are four lines connecting the
points $O_{\rm AA} (-32,6)$, $O_{\rm CC} (32,6)$, $O_{\rm FF}
(32,-6)$ and $O_{\rm PP} (-32,-6)$ in the parameter space of
$(C_{12}\gamma,C_{12}\beta)/|C_1\beta_1|$.} \label{fig1}
\end{figure}

To quantify entanglement between the two species, we numerically computed
the von Neumann entropy
$S(\hat{\rho}_1)=-\text{Tr} (\hat{\rho}_1\log_{2N+1} \hat{\rho}_1)$,
where $\hat{\rho}_1=\text{Tr}_2\hat{\rho}$ is the reduced density
matrix resulting from partial tracing of the ground-state density
matrix $\hat{\rho}$ over the basis of species 2.
The amount of entanglement is then shown as density plots
over the phase diagram of Fig. \ref{fig1}, with a legend
shown on the right,
the black (white) color refer to low (high) entanglement.

In the absence of interspecies singlet-pairing interaction $(\gamma=0)$,
the spin-dependent Hamiltonian of Eq.~(\ref{hamiltonian}) contains
only three operators commuting with each other. As a result we can use
quantum numbers $l_1$, $l_2$, $l$, and $l_z$ to quantify its
eigenstate $|l_1,l_2,l,l_z\rangle$, with $\langle\hat{L}_1^2\rangle=l_1(l_1+1)$,
$\langle\hat{L}_2^2\rangle=l_2(l_2+1)$, $\langle\hat{L}^2\rangle=l(l+1)$
and $\langle\hat{L}_z\rangle=l_z$. In the ground states, we have
$l=l_1+l_2$ for ferromagnetic interspecies spin-exchange interaction $(\beta<0)$;
and $l=|l_1-l_2|$ for anti-ferromagnetic interspecies spin-exchange interaction
$(\beta>0)$, while the value of $l_1$ and $l_2$ are determined by the
three interaction parameters.

\begin{figure}[H]
\centering
\includegraphics[width=3.4in]{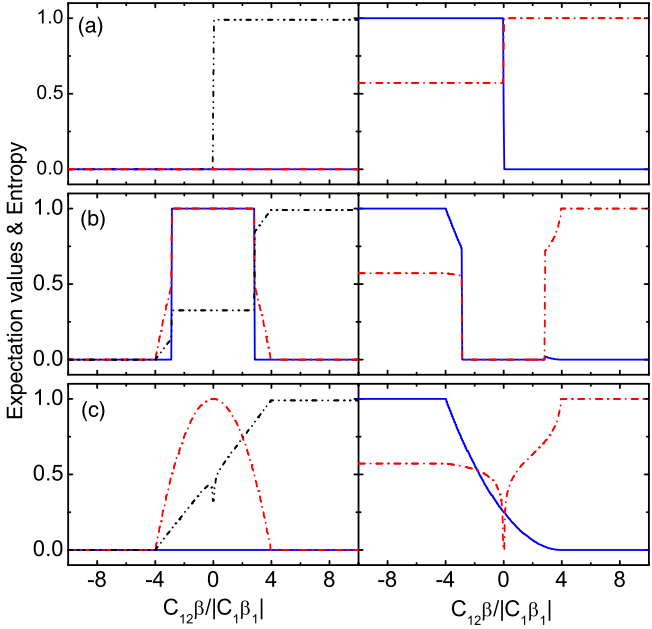}
\caption{(Color online). LEFT: The ground-state normalized expectation values of intra-
and interspecies singlet-pairing number operators $\hat{A}^{(2)\dag}_0\hat{A}^{(2)}_0/N(N+1)$,
$\hat{B}^{(2)\dag}_0\hat{B}^{(2)}_0/N(N+1)$, $\hat{\Gamma}^{(1,1)\dag}_0\hat{\Gamma}^{(1,1)}_0/N(N+2)$,
denoted by blue solid, red dash-dot, and black dash-dot-dot lines, respectively.
RIGHT: The normalized total spin angular momentum $\hat{L}^2/2N(2N+1)$ and von Neumann entropy
of the ground state, denoted by blue solid and red dash-dot lines, respectively.
The three subplots denote zero interspecies singlet-pairing interaction $(\gamma=0)$ and
fixed intra-species spin-exchange interaction
parameters
$(C_1\beta_1, C_2\beta_2)/|C_1\beta_1|$ $=$: (a) ($-1$,$-2$) ; (b)
($1,2$); and (c) ($-1,2$). } \label{fig2}
\end{figure}

In Fig. \ref{fig2}, we display the expectation values of the
intra- and interspecies single-pairing number operators in the ground state,
divided by their corresponding maximum values shown in Tab. \ref{tabl}.
In addition, we show the total spin angular momentum and
the von Neumann entropy.
First, when $(\beta_1<0, \beta_2<0)$, irrespective of the
interspecies spin-exchange interaction,
atoms in the same species will not pair into a singlet,
but atoms in different species will pair into singlets with
$\langle \Gamma_0^{(1,1)\dag}\Gamma_0^{(1,1)}\rangle$ close to
reaching its maximum values $N(N+2)$ when the
interspecies spin-exchange interaction is anti-ferromagnetic.
The corresponding total spin angular momentum $\hat{L}^2$ is equal to
its maximum value $2N(2N+1)$ with a relatively low von Neumann entropy
when the interspecies spin-exchange interaction is ferromagnetic.
The total spin angular momentum $\hat{L}^2$
is equal to $0$ with the von Neumann entropy $S(\hat{\rho}_1)=1$
for anti-ferromagnetic
interspecies spin-exchange interaction.
Second, when $(\beta_1>0,\beta_2>0)$, in the two
limits of large ferromagnetic or anti-ferromagnetic interspecies
spin-exchange interaction, the ground state is the same as above
in the previous case. In the other limit with low
interspecies spin-exchange interaction, atoms in the same species
tend to pair into singlets giving rise to no entanglement between the two species.
This implies the ground state can be written as a product
state: $Z^{-1/2}(\hat{A}_0^{(2)\dag})^{N/2}(\hat{B}_0^{(2)\dag})^{N/2}|\rm vac\rangle$.
For the remaining phase, we call it the MM phase,
which can show higher (lower) entanglement compared to the FF phase
between the two species when $\beta>0$ ($\beta<0$).
For the final case when $(\beta_1<0,\beta_2>0)$,
the ground state show similar properties as that in the second case.

\begin{figure*}[t]
\centering
\includegraphics[width=6.4in]{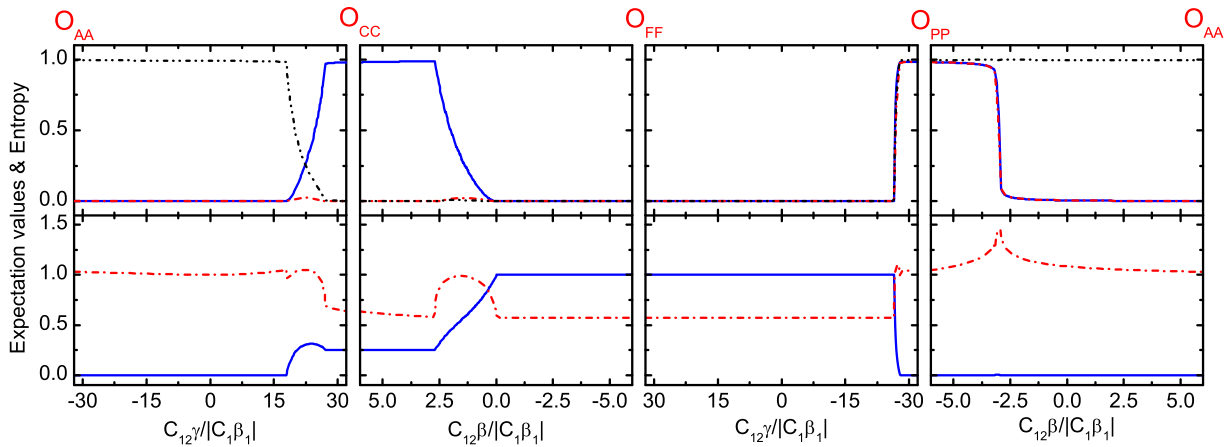}
\caption{(Color online). TOP: The ground-state normalized
expectation values of intra- and interspecies singlet-pairing number
operators $\hat{A}^{(2)\dag}_0\hat{A}^{(2)}_0/N(N+1)$,
$\hat{B}^{(2)\dag}_0\hat{B}^{(2)}_0/N(N+1)$,
$\hat{\Gamma}^{(1,1)\dag}_0\hat{\Gamma}^{(1,1)}_0/N(N+2)$, denoted
by blue solid, red dash-dot, and black dash-dot-dot lines,
respectively. BOTTOM: The normalized total spin angular momentum
$\hat{L}^2/2N(2N+1)$ and Von Neumann entropy of the ground state,
denoted by blue solid and red dash-dot lines, respectively. From
left to right, we illustrate the corresponding expectation values
and Von Neumann entropy along four direct lines connecting four
points $O_{\rm AA} (-32,6)$, $O_{\rm CC} (32,6)$, $O_{\rm FF}
(32,-6)$ and $O_{\rm PP} (-32, -6)$ in the parameter space of
$(C_{12}\beta,C_{12}\gamma)/|C_1\beta_1|$. The intraspecies
spin-exchange interactions are fixed at $(C_1\beta_1,
C_2\beta_2)/|C_1\beta_1| =(-1,-2)$. The four lines are marked as red
dashed lines in the Fig. \ref{fig1}(a).} \label{fig3}
\end{figure*}
The most attractive phase when $\gamma=0$ is
the entangled ground state denoted by $\psi_{\rm AA}^{00}$ \cite{xu2010},
\begin{eqnarray}
  \psi_{\rm AA}^{00}=\frac{1}{\sqrt{2N+1}}\sum\limits_{m=-N}^N
  (-)^{N-m}|N,m\rangle_1\otimes|N,-m\rangle_2,
  \label{psiAA00}
\end{eqnarray}
which show high entanglement with $S(\hat{\rho}_1)=1$ between the
two species. As demonstrated in Fig. \ref{fig1} by numerical
calculations, however, there remain other phases which show larger
entanglement between the two species. This shows that the state
$\psi_{\rm AA}^{00}$ is not a maximal entangled state, in contrast
to previously studied case of two pseudo spin-1/2 condensates
\cite{shi2006}. This is not a surprise \cite{you2002}. Due to the
redundant degrees of freedom in the spin-1 case, the total spin
angular momentum of each species can take other values besides the
largest value of $N$. To demonstrate the entanglement between the
two species, we show their corresponding expectation values and von
Neumann entropy along four lines connecting the points $O_{\rm AA}
(-32,6)$, $O_{\rm CC} (32,6)$, $O_{\rm FF} (32,-6)$ and $O_{\rm PP}
(-32, -6)$ in the parameter space of
$(C_{12}\gamma,C_{12}\beta)/|C_1\beta_1|$. The four lines are marked
as red dashed lines in the Fig. \ref{fig1}(a).

In Fig. \ref{fig3}, we illustrate the ground-state properties of
two ferromagnetic condensates with intraspecies spin-exchange interactions
at $(C_1\beta_1,C_2\beta_2)/|C_1\beta_1|=(-1,-2)$.
First of all, we consider the AA phase.
When $\gamma=0$, the spin-dependent Hamiltonian contains three operators
commuting with each other, and the ground state can be expressed as
$\psi_{\rm AA}^{00}$ for large enough anti-ferromagnetic interspecies
spin-exchange interaction, and show high entanglement between the
two species. When $\gamma\ne0$, although the fourth term of the spin-dependent
Hamiltonian does not commute with the other three, we find that the ground state
not only show similar expectation values of the operators, it
also contain similar entanglement between the two species, over a large area
in the phase diagram demonstrated in Fig. \ref{fig1}(a).
For $\gamma<0$, irrespective of its value, the ground state
falls into the AA phase. For $\gamma>0$, the ground state
is still classified as the AA phase, as long as $C_{12}\gamma$ does not exceed a critical value,
which increases in proportion to the interspecies spin-exchange interaction parameter
$C_{12}\beta$. In the first column of Fig. \ref{fig3},
we evaluate the properties of the AA phase, where the ground-state
expectation value of intra- and interspecies singlet-pairing number operators
are close to $0$, $0$, and $N(N+2)$, respectively.
The total spin angular momentum is exactly equal to 0,
and the von Neumann entropy is close to 1.

\begin{figure*}
\centering
\includegraphics[width=6.4in]{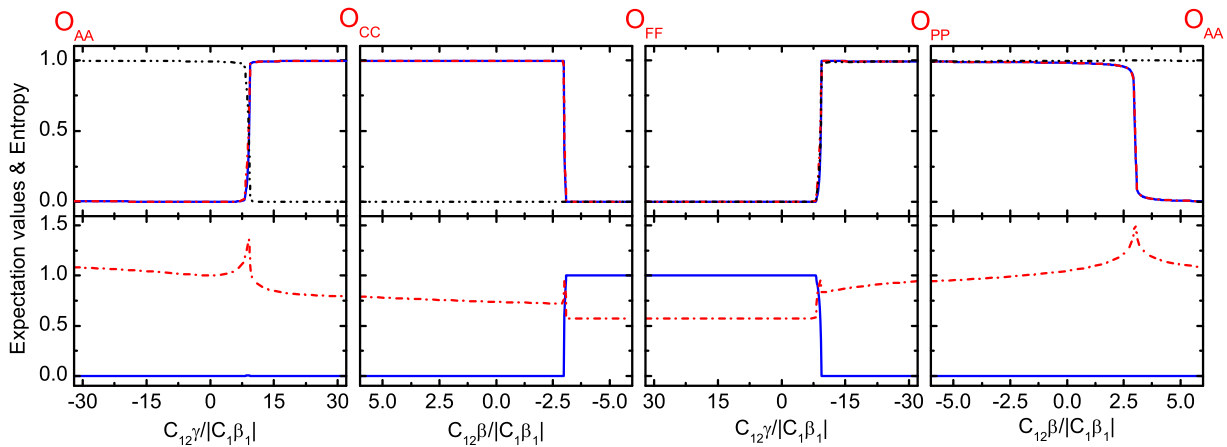}
\caption{(Color online). The same as that in Fig. \ref{fig3}, but
with different intraspecies spin-exchange interactions fixed at
$(C_1\beta_1, C_2\beta_2)/|C_1\beta_1| =(1,2)$.}
\label{fig4}
\end{figure*}

\begin{figure*}
\centering
\includegraphics[width=6.4in]{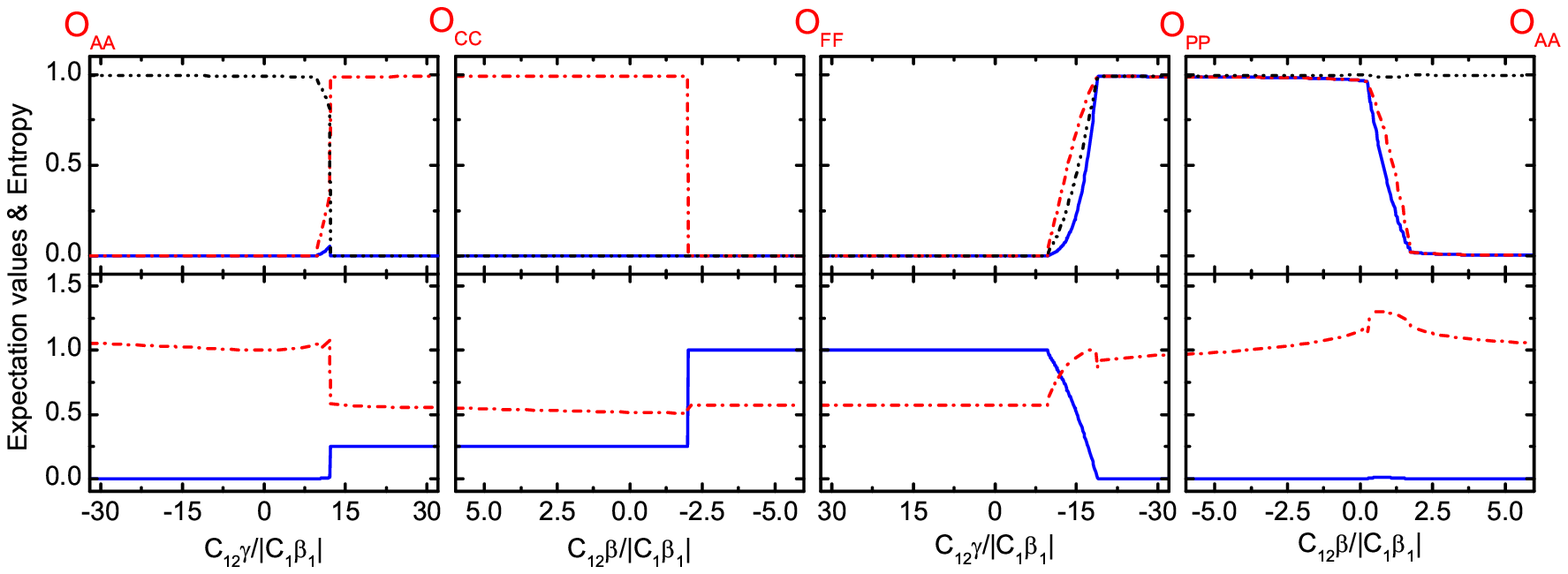}
\caption{(Color online). The same as that in Fig. \ref{fig3}, but
with different intraspecies spin-exchange interactions, which are fixed at
$(C_1\beta_1, C_2\beta_2)/|C_1\beta_1| =(-1,2)$.}
\label{fig5}
\end{figure*}

When interspecies singlet-pairing interaction exceeds a critical
value, the ground state changes into the MM phase.
As long as $C_{12}\gamma>0$, it tries to decrease the interspecies
singlet-pairing interaction.
In the first column of Fig. \ref{fig3},
we follow the line of $O_{\rm AA}O_{\rm CC}$
and illustrate the phase transition from the AA phase into the MM phase.
In the MM phase, as long as we increase $C_{12}\gamma$ accordingly,
atoms in different species will continuously avoid to pair into singlets,
while atoms in species 1 will try to pair into singlets.
Atoms in species 2 first try to pair into singlets and then avoid to pair.
Meanwhile, the total spin angular momentum first
increases and then decreases. For a relatively large area of the MM phase,
we find the two species show high entanglement compared to the AA phase.

As interspecies singlet-pairing interaction is increased, the ground state
will fall into the CC1 phase, where its total spin angular momentum $\hat{L}^2$
will be equal to $N(N+1)$, and atoms in the species 1 (2)
will pair (not pair) into singlets. At the same time the expectation value of
$\Gamma_0^{(1,1)\dag}\Gamma_0^{(1,1)}$ will be near to its maximum $N(N+2)$.
The ground-state von Neumann entropy in the CC1 phase remains at a low value.

Going along the line $O_{\rm CC}O_{\rm FF}$, with
decreasing interspecies spin-exchange interaction,
the ground state changes from the CC1 phase,
to the MM phase, and finally to the FF phase.
Atoms of species 1 will become unpaired continuously,
while the total spin angular momentum increases from $N(N+1)$
in the CC1 phase to its maximum $2N(2N+1)$ in the FF phase.
The ground-state entanglement between the two species
in the CC1 and the FF phase are almost at the same level.
While in the MM phase, the entropy first increases to near
1 and then decreases.

We then follow the line $O_{\rm FF}O_{\rm PP}$.
With decreasing interspecies singlet-pairing interaction,
the ground state covers the FF, MM, and PP phases successively.
In the PP phase, atoms in the same/different species will all
try to pair into singlets. The ground-state expectations values
for intra- and interspecies singlet-pairing number operators
will reach close to their corresponding maximum,
meanwhile the two species show higher entanglement.
In the MM phase, the expectations values
for operators or the von Neumann entropy change
continuously to connect the FF phase and the PP phase.

Lastly, we consider the line $O_{\rm PP}O_{\rm AA}$.
As long as the interspecies spin-exchange interaction increases,
the ground state will be changed continuously from the PP to the MM,
and to the AA phase. From the fourth column of Fig. \ref{fig3},
we find that in the whole line of $O_{\rm PP}O_{\rm AA}$,
the two species show a relatively high entanglement with
the von Neumann entropy $S(\hat{\rho}_1)$ remaining higher than 1.
Especially so in the MM phase, where the highest entropy reaches
1.4568, which is close to the maximum entropy
of $\log_{2N+1}((N+1)(N+2)/2)\simeq 1.6116$.

In Figs. \ref{fig4} and \ref{fig5}, we illustrate the
two other cases with intraspecies spin-exchange interactions
$(C_1\beta_1,C_2\beta_2)/|C_1\beta_1|$ fixed respectively at $(1,2)$
and $(-1,2)$. We find that the ground-state show
similar properties to that in the two ferromagnetic condensates
shown in Fig. \ref{fig3}. The only difference is for the CC2 or CC3 phase.
In the CC2 phase, the ground state expectation values for
both intra- and interspecies
singlet-pairing number operators are close to their corresponding
maximum $N(N+1)$, $N(N+1)$, and $N(N+2)$.
While in the CC3 phase, they are close to $0$, $N(N+1)$, and $N(N+2)$, respectively.
The total spin angular momentum of the ground state
is equal to $0$ in the CC2 phase, and $N(N+1)$ in the CC3 phase.
Meanwhile, in the CC2 phase the von Neumann entropy is less than 1 but larger than
that in the FF phase, while in the CC3 phase, it is close to the value in the FF phase.

Before conclusion, we hope to stress that the maximal entangled state in
this system is given by
\begin{eqnarray}
  \psi_{\rm ME}=Z^{-1/2}\left(\hat{\Gamma}_0^{(1,1)\dag}\right)^N|\rm vac\rangle.
  \label{maximales}
\end{eqnarray}
It is the eigenstate or the ground state (if $\gamma<0$) of the $\gamma$-term
in the Hamiltonian of the Eq.~(\ref{hamiltonian}),
which means the maximal entangled state $\psi_{\rm ME}$ is
the eigenstate of two spin-1 condensates with only interspecies
spin-singlet pairing interaction ($\beta_1=\beta_2=\beta=0$ and $\gamma\ne0$),
with the corresponding eigenvalue $C_{12}\gamma N(N+2)/6$ \cite{mistake}.

\section{Conclusion}

In conclusion, we have studied the ground-state phase diagram for a
binary mixture of two spin-1 condensates more carefully, going
beyond the MF approximation.
When there exists no interspecies singlet-pairing interaction,
the spin-dependent Hamiltonian contains three operators commuting with each other.
In this special case, the most interesting phase is the AA phase, where
two species show high entanglement. When interspecies singlet-pairing interaction
is turned on, the added operators do not commute with the previous three ones,
which forbids us from obtaining exact eigenstates analytically for the model spin system.
In this study, we perform full quantum diagonalization to
find the ground states numerically.
To quantify the ground states, we work out the building blocks
to construct the maximum spin states, which can be used rightfully
to discuss entanglement scales between the two species.
We have evaluated the associated ground-state von Neumann entropy.
After detail calculations, we find that the AA phase can persist for large
areas of the parameter space for interspecies spin-exchange and singlet-pairing
interactions. In addition, there is another interesting phase: the PP phase,
which show similar level of entanglement between the two species.
What's more, we find that AA phase is not the maximum entangled state.
The ground state with highest entanglement we found lies in the MM phase.

\section{Acknowledgments}

This work is supported by NSF of China under Grants No.~11004116, No.~10974112
and No.~91121005 , NKBRSF of China, and the research program 2010THZO
of Tsinghua University.


\begin{thebibliography}{10}
\bibitem{numerical}
  The total Hilbert space for zero magnetization $l_z=0$ of the case of
  $N_1=N_2=100$ is 174301. Here we use ARPACK pakage to do numerical
  diagonalization.
\bibitem{mistake}
  We have previously miscalculated the operator $\hat{\mathcal{S}}_z$, which is equal
  to $(\hat{N}_1+\hat{N}_2+3)/2$ not $(\hat{N}_1+\hat{N}_2)/2+3$ in the Ref. \cite{xu2010}.
  As a result, $\mathcal{S}=(n_0+3)/2$ $(n_0=0,1,2,\dots)$, $\min(\mathcal{S})=3/2$
  and $\min(\mathcal{S}_z)=3/2$. Also the Eq. (13) in the Ref. \cite{xu2010} need to be changed
  to $ E=\frac{1}{4}C_{12}\beta\, l(l+1)+\frac{1}{6}C_{12}\gamma\, n_s(n_s+n_0+2)
  -\frac{1}{2}C_{12}\beta(N^{(1)}+N^{(2)})$.
\end{thebibliography}
\end{document}